\documentclass{llncs}
\usepackage{color}
\usepackage{graphics}
\usepackage{epsfig}

\newcommand{\MB}{\[\begin{array}{lllll}}
\newcommand{\ME}{\end{array}\]}
\newcommand{\oomit}[1]{}
\newcommand{\HH}{\mbox{\^{}}}

\newcommand{\noop}[1]{}
\newcommand{\split}{\textit{ split }}
\newcommand{\vin}[1] { v^{in}_{#1} }
\newcommand{\vout}[1] { v^{out}_{#1} }
\newcommand{\locationtree}{\textit{ location tree }}
\newcommand{\existss}{\textit{ exists }}
\newcommand{\exist}{\textit{ exist }}
\newcommand{\feasible}{\textit{ feasible }}
\newcommand{\infeasible}{\textit{ infeasible }}

\begin{document}

\title{Numerical Simulation guided Lazy Abstraction Refinement for  Nonlinear Hybrid Automata}
\author{ Sumit Kumar Jha\inst{1}}
\institute{ Computer Science Department, Carnegie Mellon
University\\ 5000 Forbes Avenue, Pittsburgh, PA15213,
USA\\jha+@cs.cmu.edu \\
}

\maketitle

\begin{abstract}
This draft suggests a new counterexample guided abstraction
refinement (CEGAR) framework that uses the combination of numerical
simulation for nonlinear differential equations with linear
programming for linear hybrid automata (LHA) to perform reachability
analysis on nonlinear hybrid automata. A notion of $\epsilon-$
structural robustness is also introduced which allows the algorithm
to validate counterexamples using numerical simulations.

\end{abstract}

\section{Introduction}

The model checking of hybrid automata remains a challenge
 and the existing tools \cite{henzinger97hytech,DBLP:conf/hybrid/Frehse05}
do not scale up to the needs of the industry. Because of the well
known fundamental undecidability results \cite{thatom}, the model
checking of general hybrid automata often proceeds by building
successive tighter approximations to these hybrid automata in a
relatively easy-to-analyze fragment of hybrid automata like Linear
Hybrid Automata \cite{hothesis}. Theoretical results about the
asymptotic completeness of this approximation procedure form the
backbone of such a  strategy behind the model checking of nonlinear
hybrid automata.

There has been considerable interest in applying Counterexample
Guided Abstraction Refinement (CEGAR), which works so well with
discrete systems, to the problem of hybrid system verification
\cite{alur03counterexample}. There has also been some exploration of
using fragments instead of counterexamples during abstraction
refinement \cite{DBLP:conf/hybrid/FehnkerCJK05} and the application
of CEGAR specifically to LHA \cite{jha2007}. However, our ongoing
work makes the following new contributions to the abstraction
refinement based analysis of hybrid systems.

\begin{itemize}

\item We address the problem of abstraction refinement for nonlinear
hybrid automata and use CEGAR to construct successively refined LHA
approximations. Our refinement is lazy and hence, refines some parts
of the state space more finely than others.

\item  We use the distance between a feasible path in the abstract linear hybrid automata
 and the numerically simulated trajectory in the
nonlinear hybrid automata to refine those locations in the LHA that
do not faithfully represent the behavior of the nonlinear hybrid
automata.

\item We define a structural notion of robustness and use it to
 present a counterexample validation algorithm (for a rich
class of nonlinear hybrid automata) using linear programming
\cite{jha2006}. Hence, it is possible to detect reachability of a
bad state even before the abstraction refinement loop terminates.

\end{itemize}

\section{Background on LP based path feasibility analysis of LHA}

Informally, a linear hybrid automaton is a conventional automaton
extended with a set of continuous variables. The states of the
automaton called {\it locations} are annotated with a change rate
for each continuous variable such as $\dot{x}=[a,b]$ ($x$ is a
variable, and $[a,b]$ is a rational interval), and the transitions
of the automaton are labeled with constraints on the variables such
as $a\leq\sum^m_{i=0}c_ix_i\leq b$ and /or with reset actions such
as $x:=c$ ($x_i$ and $x$ are variables, $a$, $b$, and $c_i$ are real
numbers). Such linear hybrid automata are essentially equivalent to
the definition given in \cite{thatom}. It is known that this
subclass of linear hybrid automata are sufficiently expressive to
allow asymptotic completeness of the abstraction process for a
general hybrid automata. `` A restricted form of linear phase
portrait approximations are asymptotically complete, namely, when
all automaton constraints are over-approximated using independent,
rational lower and upper bounds on the values and derivatives of
each variable" \cite{alur95algorithmic}.For simplicity, we suppose
that in any linear hybrid automaton considered in this paper, there
is just one initial location with no initial conditions and no
transitions to the initial location (we assume that each variable
with an initial value is reset to the initial value by the
transitions from the initial location).

\begin{definition} \rm A linear hybrid automaton is a tuple $H=(X,V,E,v_I,\alpha,\beta)$, where

\begin{itemize}
\item $X$ is a finite set of real-valued variables.
\item $V$ is a finite set of {\it locations}.
%
\item $E$ is {\it transition} relation whose elements are of the
form $(v,\phi,\psi,v')$ where $v,v'$ are in $V$, $\phi$ is a set of
{\it guards or variable constraints} of the form $a\leq
\sum_{i=0}^mc_ix_i\leq b$, and $\psi$ is a set of {\it reset
actions} of the form $x:=c$ where $x_i\in X\ (0\leq i\leq m)$, $x\in
X$, $a,b$ and $c_i\ (0\leq i\leq m)$ are real numbers, and $a$ and
$b$ may be $\infty$.
\item $v_I$ is an {\it initial} location.
\item $\alpha$ is a labeling function which maps each location in
$V-\{v_I\}$ to a {\it state invariant} which is a set of variable
constraints of the form $a\leq\sum^m_{i=0}c_ix_i\leq b$ where
$x_i\in X$ $(0\leq i\leq m)$, $y\in X$, $a,b$, and $c_i\ (0\leq
i\leq m)$ are real numbers, $a$ and $b$ may be $\infty$.
\item $\beta$ is a labeling function which maps each location in
$V-\{v_I\}$ to a set of {\it change rates} which are of the form
$\stackrel{.}{x}=[a,b]$ where $x\in X$, and $a, b$ are rational
  numbers ($a\leq b$). For any location $v$, for any $x\in X$, there is one and only one
change rate definition $\stackrel{.}{x}=[a,b]\in \beta(v)$.
\end{itemize}
\qed
\end{definition}

For a linear hybrid automaton $H=(X,V,E,v_I,\alpha,\beta)$ , a {\it
path segment} is a sequence of locations \MB
v_1\stackrel{(\phi_1,\psi_1)}{\longrightarrow}
v_2\stackrel{(\phi_2,\psi_2)}{\longrightarrow}\dots\stackrel{(\phi_{n-1},\psi_{n-1})}{\longrightarrow}v_{n}\ME
which satisfies $(v_i,\phi_i,\psi_i,v_{i+1})\in E$ for each $i\
(1\leq i\leq n-1)$. A {\it path} in $H$ is a path segment starting
at $v_I$. The behavior of linear hybrid automata can be represented
by {\it timed sequences}. Any timed sequence is of the form
$(v_1,t_1)\HH(v_2,t_2)\HH\ldots\HH (v_n,t_n)$, where $v_i\ (1\leq
i\leq n)$ is a location and $t_i\ (1\leq i\leq n)$ is a nonnegative
real number, which represents a behavior of an automaton that the
system starts at the initial location and changes to the location
$v_1$, stays there for $t_1$ time units, then changes to the
location $v_2$ and stays in $v_2$ for $t_2$ time units, and so on.

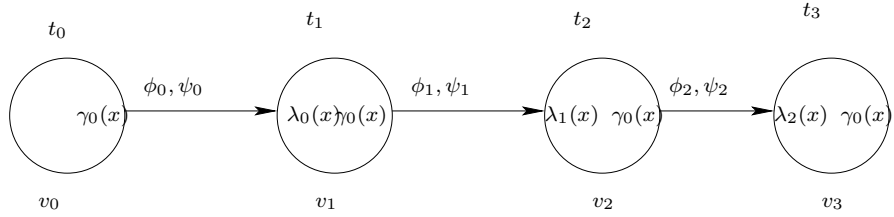
\begin{figure}

\input{lpfeb2006_path.pstex_t}
\end{figure}

\begin{definition} ~\cite{jha2006} \rm For a linear hybrid automaton
$H=(X,V,E,v_I,\alpha,\beta)$, a timed sequence $(v_1,t_1)
\mbox{\^{}} (v_2,t_2) \mbox{\^{}} \dots \mbox{\^{}} (v_n,t_n)$
represents a behavior of $H$ if the following condition is
satisfied:
\begin{itemize}
\item there is a path in $H$ of the form \MB
v_0\stackrel{(\phi_0,\psi_0)}{\longrightarrow}
v_1\stackrel{(\phi_1,\psi_1)}{\longrightarrow}\dots\stackrel{(\phi_{n-1},\psi_{n-1})}{\longrightarrow}v_{n}\,;
\ME

\item $t_1,t_2,\dots ,t_n$ satisfy all the variable constraints in
$\phi_i\ (1\leq i\leq n-1)$, i.e. for each variable constraint
$a\leq c_0x_0+c_1x_1+\dots+c_mx_m\leq b$ in ${\phi}_i$, \MB
\delta_k\leq \gamma_i(x_k)\leq\delta_k'\ \mbox{for any $k\ (0\leq
k\leq m)$, and}\\ a\leq
c_0\gamma_i(x_0)+c_1\gamma_i(x_1)+\dots+c_m\gamma_i(x_m)\leq b\ME
 where $\gamma_i(x_k)\ (0\leq k\leq m)$ represents the value of the variable
 $x_k$ when the automaton stay at $v_i$ with the delay
 $t_i$; and, similarly,

\item  $t_1,t_2,\dots ,t_m, \gamma_i(x_k), \lambda_i(x_k)$ satisfy
the state invariant for each location $v_i$ $(1\leq i\leq n)$, where
$\gamma_i(x_k)\ (0\leq k\leq m)$ represents the value of the
variable  $x_k$ when the automaton stay at $v_i$ with the delay
 $t_i$, and $\lambda_i(x_k)\ (0\leq k\leq m)$ represents the value
 of the variable $x_k$ after leaving state $v_i$ and after the reset
 conditions have been applied.

\end{itemize}
\end{definition}

Now, we use linear programming to test the feasibility of a single
path for the reachability analysis of linear hybrid automata. Let
$H=(X,V,E,v_I,\alpha,\beta)$ be a linear hybrid automaton, , and
$\rho$ be a path in $H$ of the form \MB
v_0\stackrel{(\phi_0,\psi_0)}{\longrightarrow}
v_1\stackrel{(\phi_1,\psi_1)}{\longrightarrow}\dots\stackrel{(\phi_{n-1},\psi_{n-1})}{\longrightarrow}v_{n}
\ME where $v_n=v$. For any timed sequence of the form
$(v_1,t_1)\HH(v_2,t_2)\HH\dots\HH(v_n,t_n)$, if $\rho$ is feasible,
then the following condition must hold:

\begin{itemize}

\item $t_1,t_2,\dots,t_n$ satisfy all the variable constraints in
$\phi_i (0\leq i\leq n)$, and

\item $t_1,t_2,\dots,t_n$ satisfy all the variable constraints in
$\alpha(v_i)\ (1\leq i \leq n)$,

\end{itemize}

which form a group of linear inequalities on
$t_1,t_2,\dots,t_n,\gamma_i ({x_k}),\lambda_i ({x_k})$ (see
Definition 2), denoted by $\Theta(\rho)$ or $LP_{\rho}
(t_i,\gamma_i({x_k}), \lambda_i({x_k}) ) $. It follows that we can
check if $\rho$ is a feasible path by checking if the group
$\Theta(\rho)$ (or $LP_{\rho} (t_i,\gamma_i({x_k}), \lambda_i({x_k})
) $) of linear inequalities has a solution, which can be solved by
linear programming ~\cite{jha2006}.

\section{Background on Abstraction of Affine dynamics by LHA}

Given a general hybrid system $H=(X,V,E,v_I,\alpha,\beta)$ where $X$
is a finite set of real-valued variables, $V$ is a finite set of
{\it locations}, $E$ is {\it transition} relation whose elements are
of the form $(v,\phi,\psi,v')$ where $v,v'$ are in $V$, $\phi$ is a
set of {\it guards or variable constraints} of the form $a\leq
\sum_{i=0}^mc_ix_i\leq b$, and $\psi$ is a set of {\it reset
actions} of the form $x:=c$ where $x_i\in X\ (0\leq i\leq m)$, $x\in
X$, $a,b$ and $c_i\ (0\leq i\leq m)$ are real numbers, and $a$ and
$b$ may be $\infty$, $v_I$ is an {\it initial} location, $\alpha$ is
a labeling function which maps each location in $V-\{v_I\}$ to a
{\it state invariant} which is a set of variable constraints of the
form $a\leq\sum^m_{i=0}c_ix_i\leq b$ where $x_i\in X$ $(0\leq i\leq
m)$, $y\in X$, $a,b$, and $c_i\ (0\leq i\leq m)$ are real numbers,
$a$ and $b$ may be $\infty$, $\beta$ is a labeling function which
maps each location in $V-\{v_I\}$ to a set of {\it change rates}
which are of the form $\stackrel{.}{x_i}=f(x_0,x_1,\dots,x_m,\dot
x_0,\dot x_1,\dots,\dot x_m, a_0,a_1 \dots a_m)$ where $x_0,x_1
\dots x_m \in X$, and $a_0, a_1 \dots, a_n$ are real numbers). For
any location $v$, for any $x\in X$, there is one and only one change
rate definition.

We construct (in a fashion similar to
\cite{henzinger97hytech,DBLP:conf/hybrid/Frehse05}) a linear hybrid
automata $H_{a}$ which is an over-approximate abstraction of the
general hybrid automata H. We define an operator \split which
divides a location into smaller locations and use this to divide
each location $v_i$ of the original hybrid automaton.

\begin{definition} Let v be a location  and  $x_i \in X$ be a variable in
a hybrid automata $H = (X,V,E,v_i,\alpha,\beta)$ . Suppose $In_v =
\{ \vin{0}, \vin{1} \dots \vin{n} \}$ be the locations from which
there is a transition into the location v and $Out_v = \{ \vout{0},
\vout{1} \dots \vout{m} \} $ be the locations to which there is a
transition from v. Also, $C$ be a set of linear constraints on $X$.
Then, the operator \split $(H,v,C)$ constructs a new hybrid automata
$ H' = ( X , V' , E' , v_i, \alpha', \beta') $, where

\begin{itemize}

\item $ V' = V \cup \{ v' , v'' \}  \setminus {v}  $

\item $ \alpha' = \alpha \cup \{\alpha(v'), \alpha(v'') \} \setminus
\{ \alpha(v) \}$, where

\begin{itemize}

\item $\alpha(v') = \alpha(v) \cup \{ C \}$

\item $\alpha(v'') = \alpha(v) \cup \{ \neg C \}$

\end{itemize}

\item $ E' =  E$  $\setminus $  $($ $\{ (v_{in}, \phi_{v_{in},v},
\psi_{v_{in},v },  v) | v_{in} \in  In_v\} \cup \{ ( v, \phi_{v ,
v_{out}}, \psi_{v , v_{out}}, v_{out} ) | v_{out} \in Out_v \} $ $ )
$ $ \cup $  $ ( $   $ \{ (v_{in}, \phi_{v_{in},v}, \psi_{v_{in},v},
v') | v_{in} \in In_v\} \cup \{ v', \phi_{v , v_{out}}, \psi_{v ,
v_{out}}, v_{out} | v_{out} \in Out_v \} \cup \{
(v_{in},\phi_{v_{in},v}, \psi_{v_{in},v}, v'') | v_{in} \in In_v \}
\cup \{ v'', \phi_{v , v_{out}}, \psi_{v , v_{out}}, v_{out} |
v_{out} \in Out_v \} $ $)$  $  \cup  $ $\{ (v', \alpha(v',v'') =
C,\{ \}, v'') $ $,$  $ (v'',\alpha(v'',v') = \neg C,\{ \},v') \}$\\

\item $\beta' = \beta \cup \{ \beta(v') , \beta(v'') \} \setminus
\beta(v)$, where

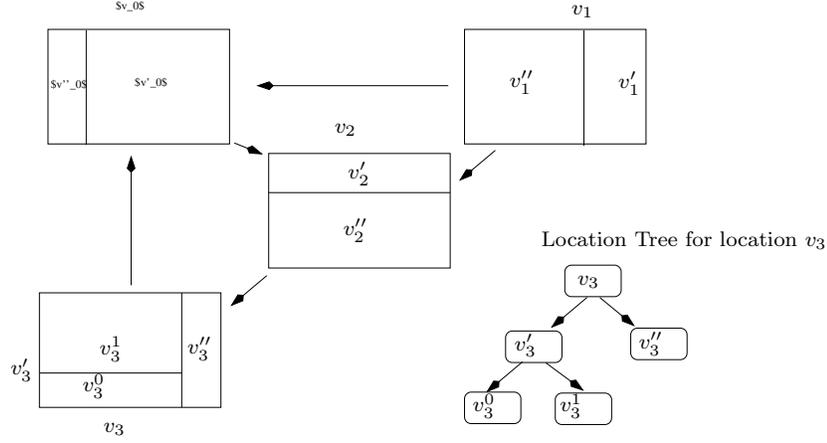
\begin{itemize}

\item $\beta(v') = \beta(v)$

\item $\beta(v'') = \beta(v)$

\end{itemize}

\end{itemize}

\end{definition}

\begin{figure}

\input{lpfeb2006_split.pstex_t}

\end{figure}

We call the new locations $v'$ and $v''$ as the children of $v$. In
particular, $v' = child(v,C)$ and $v'' = child(v,\neg C)$, where C
is the set of linear constraints used to split v. We also call $v$
as the parent of $v'$ and $v''$. Thus, the \split operator naturally
defines a tree of locations that we call the \locationtree , where
the children location are formed by splitting the parent location.

\begin{definition}
LHA-approximation to a general hybrid automata: Given a general
hybrid automata $H = (X,V,E,v_i,\alpha,\beta)$, $H_a =
(X,V_a,E_a,V_{i_a},\alpha_a,\beta_a)$ is a LHA-approximation
\textit{iff}

\begin{itemize}

\item There exists a hybrid automata $H' =
(X,V',E',V'_i,\alpha',\beta')$, where $H' = split^n(H)$.

\item $V_a = V', E_a = E', V'_i = V_{i_a}, \alpha' = \alpha_a$

\item $\forall v \in V_a, \beta_a(v) \supset \beta'_v$ and
if $c \in \beta_a(v)$, then $c$ is of the form $ x:=[a,b] $, where
$a,b \in \mathcal{R}$.

\end{itemize}

\end{definition}

\section{Definitions}

The linear hybrid automaton $H_a$ is an over-approximate
approximation of the general hybrid automata $H$. A path $\rho = \{
v_0, v_1, \dots v_m \}$, where $v_i \in V$ is said to \exist in
$H_a$ if $(v_i,v_{i+1}) \in E, 0 \leq i < m$.

Consider a path $\rho = \{ v_0, v_1, \dots v_m \}$ that \existss in
the abstract linear hybrid automata model $H_a$ and let $LP_{\rho}
(t_i,\gamma_i({x_k}), \lambda_i({x_k}) ) $ be the linear program
corresponding to the path. If the linear program $LP_{\rho}$ has a
feasible solution, then the path is said to be \feasible in the
abstract model i.e. the linear hybrid automaton; otherwise it is
said to be \infeasible .

\begin{definition}

\textit{Trace}: Given a \feasible path $\rho$ in the abstract linear
hybrid automaton model $H_a$, the feasible solution to the LP
program $LP_{\rho} (t_i,\gamma_i (x_k), \lambda_i (x_k) )$ is called
a \textit{trace} of $H_a$.

\noindent We write $trace(H_a) = < (v_0, \lambda_0 (x_0)$ $,
\lambda_0 (x_1) \dots $ $\lambda_0 (x_n),$ $ \gamma_0 (x_0),
\gamma_0 (x_1),$ $ \dots$$ \gamma_0 (x_n), t_0 ),$ $ ( v_1,
\lambda_1 (x_0) ,$ $ \lambda_1 (x_1)$ $ \dots $ $ \lambda_1 (x_n),$
$ \gamma_1 (x_0),$ $ \gamma_1 (x_1), $  $ \dots \gamma_1 (x_n), t_1
) \dots $ $\dots ( v_m, \lambda_m (x_0) , $ $\lambda_m (x_1) \dots
\lambda_m (x_n), $ $\gamma_m (x_0), \gamma_m (x_1),$ $ \dots
\gamma_m (x_n), t_m ) >$.

\end{definition}

It is known \cite{jha2006} that the trace obtained by the linear
program is a real execution trace of the over-approximate linear
hybrid automata.

\begin{definition}

\textit{Concretization of a path}: Consider a path $\rho = \{ v_0,
v_1, \dots v_m \}$ that is \feasible in the abstract linear hybrid
automata model $H_a$. Then, the \textit{concretization} of this path
in the original hybrid automata $H$ is the trace $\rho_{concrete} =
\{ v_0^{root}, v_1^{root} \dots v_m^{root} \} $, where $v_i^{root}$
is the root of the \locationtree in which $v_i$ is a leaf.

\end{definition}

As the $split$ operator forms a tree of locations in the abstract
linear hybrid automata, the root of the location tree is known and
the {\it concretization of an abstract path} is well defined.

\begin{definition}

\textit{Concretization of a trace}: Consider a trace $tr = < ( v_0,
\lambda_0 (x_0) $
 $, \lambda_0 (x_1) \dots \lambda_0 (x_n), \gamma_0
(x_0)$ $, \gamma_0 (x_1), \dots \gamma_0 (x_n), t_0 ), ( v_1,$
$\lambda_1 (x_0) , $ $\lambda_1 (x_1) \dots \lambda_1 (x_n),
\gamma_1$ $(x_0), \gamma_1 (x_1), \dots \gamma_1 (x_n), t_1 ) \dots
\dots ($ $v_m, \lambda_m (x_0) , \lambda_m (x_1) \dots \lambda_m
(x_n),$ $\gamma_m (x_0), \gamma_m (x_1), $ $\dots \gamma_m (x_n),
t_m ) >$ corresponding to the path $\rho = \{ v_0, v_1, \dots v_m
\}$ that is \feasible in the abstract linear hybrid automata model
$H_a$. Then, the \textit{concretization} of this \textit{trace} in
the original hybrid automata $H$ is the \textit{trace} $
tr_{concrete} = < ( v_0^{root}, \lambda_0 (x_0) ,$ $ \lambda_0 (x_1)
\dots \lambda_0 (x_n), \gamma_0 (x_0), \gamma_0 (x_1), \dots $
$\gamma_0 (x_n), t_0 ), $ $( v_1^{root}, \lambda_1 (x_0) , \lambda_1
(x_1)$ $ \dots \lambda_1 (x_n), $ $\gamma_1 (x_0), \gamma_1 (x_1),
\dots $ $\gamma_1 (x_n), t_1 )$ $ \dots \dots ( v_m^{root},
\lambda_m (x_0) , $ $\lambda_m (x_1)  $ $\dots \lambda_m (x_n),
\gamma_m (x_0),$ $ \gamma_m (x_1), \dots $ $\gamma_m (x_n), t_m )>$
, where $v_i^{root}$ is the root of the \locationtree of which $v_i$
is a leaf.

\end{definition}

\begin{definition}

\textit{$\epsilon$-Simulation Trajectory}: Given a valuation of X
i.e. $Val_0(X) = (x_0 = x0 ,x_1 = x1,\dots x_n=xn)$ in a location
$v$ of a general hybrid system $ H = (X,V,E,v_i,\alpha,\beta) $,
then $\tau (x0,x1, \dots xn,v,t)$ is said to be an $\epsilon - $
simulation trajectory for location $v$ with initial valuation $Val_0
(X)$ \textit{iff}

\begin{itemize}

\item $\tau(t=0) = (x0 , x1, \dots, xn)$

\item if $ f(x_0,x_1, \dots, x_n, t)$ is the  solution to the initial
value problem $( \beta(v), Val_0(X) )$, then $f(t) - \epsilon \leq
\tau(t) \leq f(t) + \epsilon$

\end{itemize}

\end{definition}

It is known that numerical techniques can solve the initial value
problem for ODEs (including non-linear ODEs) quiet efficiently.

\begin{definition}

{$\epsilon-$  Hybrid Simulation Trajectory}: Given an initial
valuation of X i.e. $Val_0(X) = (x_0 = x0 ,x_1 = x1,\dots x_n=xn)$
and a path $\rho = \{ v_0, v_1 \dots v_m \}$ in a general hybrid
system $H = (X,V,E,v_i,\alpha,\beta)$, then $\tau (x_0,x_1, \dots
x_n) = f(t)$ is said to be an $\epsilon - $ hybrid simulation
trajectory \textit{iff}

\begin{itemize}

\item $\tau(0) = Val_0(X)$ and $Val_0(X) \in \alpha(v_0) $

\item if $x_k := e \in \psi(v_i,v_{i+1})$ then $ \tau(x_k,\sum_0^i (t_i) + = e )
$ else $\tau  (x_k, \sum_0^i (t_i) + ) = \tau (x_k,\sum_0^i(t_i)-)$

\item Before executing the jump $(v_i,v_{i+1})$, $\tau  (x_k, \sum_0^i (t_i)
-)$ satisfies every precondition in $\phi(v_i,v_{i+1})$

\item Within each location $v_i$ where the timed path has spent time $t_i$,

\begin{itemize}

\item $\forall t,  t_i < t < t_{i+1},  \tau(v,\sum_0^i (t_i) + t)  $ is an
$\epsilon-$ simulation trajectory for location $v_i$ with initial
valuation $Val_0 = \tau( \sum_0^i (t_i) ) $.

\end{itemize}

\end{itemize}

\end{definition}

\begin{definition}

\textit{Guided Simulation Trajectory of the concretization of a
trace :} An $\epsilon - $ hybrid simulation trajectory $\tau$ is
said to be a Guided Simulation Trajectory of a concretized trace
$tr_{concrete}$ \textit{iff}

\begin{itemize}

\item  The initial valuation of X i.e. $Val_0(X) = (x_0 = x0 ,x_1
= x1,\dots x_n=xn)$ for the trajectory $\tau$ is the initial point
in the concretized trace $tr_{concrete}$.

\item The  $\epsilon - $ hybrid simulation trajectory $\tau$
corresponds to the path $\rho = \{ v_0, v_1 \dots v_m \}$
corresponding to $tr_{cocnrete}$

\end{itemize}

\end{definition}

\section{CEGAR based Refinement of the abstract linear hybrid automata}

The CEGAR algorithm repeatedly constructs LHA over-approximations to
the given (possibly nonlinear) hybrid system and then asks a LHA
analysis engine if the over-approximate LHA admits any
counterexample. If it does not, we are done and we report that the
original hybrid system has no counterexample either. Otherwise, we
take the reported counterexample of the over-approximate LHA and
attempt to validate it using numerical simulation. If we succeed in
validating the counterexample, we report an error that the bad state
is reachable and STOP. Otherwise, we find a location where we need
to split the nonlinear hybrid automata and then rebuild a more
precise over-approximate abstraction.

 \begin{center}
\framebox{
\begin{minipage}{11cm}

\begin{list}{}{}

\item \textbf{Algorithm for CEGAR}\\
\textit{(Input: Nonlinear Hybrid Automata A. Output: Error \/ No
error}

\begin{list}{}{}

\item 1. $A_0 = A$; i := 0; L = Universe.

\item 2. $LHA_i =$ LHA-approximation $(A_i)$

\item 3. $\mathcal{L} ({LHA}_i)$ := Language of ${LHA}_i$.
$\mathcal{L}$ represents the set of potential counterexamples in
$A_i$ (finitely expressible as a regular expression) \cite{jha2007}.

\item 4. $L = L$ $\cap$ $\mathcal{L} ({LHA}_i)$

\item 5. If $L$ is empty, report "BAD STATES NOT REACHABLE" and stop.

\item 6. Pick a counterexample $ce$ in $L$.

\item 7. Validate the counterexample $ce$ in the original hybrid automata $A$.

\item 8. If $ce$ is validated in A, stop and report that ERROR STATE IS
REACHABLE.

\item 9. Compute a refinement operator $split$, and $A_{i+1} = split
(A_{i})$. Also, compute $ L =  split(L)  $.

\item 10. i := i + 1

\item 11.   Loop to Step 2.

\end{list}

\end{list}

\end{minipage}

}
\end{center}

\subsection{Counterexample Validation and Structural Robustness}

Let $v_0,\gamma_0 (x_0), \gamma_0 (x_1) \dots \gamma_0 (x_n), t_0$
be the initial point in a concretized trace $tr_{concrete}$ for the
abstraction i.e. the linear hybrid system $H_a$ corresponding to the
general hybrid system $H$. Let $\tau_{(v_0,x_0,x_1, \dots x_n)}$ be
the $\epsilon-$ Hybrid Simulation Trajectory starting from this
initial point.

\begin{definition}

$\epsilon-$ Structurally Robust Hybrid System: A hybrid system $H =
(X,V,E,v_i,\alpha,\beta)$ is said to be structurally robust
\textit{iff}

\begin{itemize}

\item $\forall i, (v_i, \phi,\psi,v_{i+1}) \in E $, every constraint
$c$ in $\phi$ is satisfied by at least a dense set of size
$\epsilon$ i.e. If $S = \{ Val = ({x_0,x_1, \dots x_n}) | Val$
satisfies $c \}$, then $max_{a \in S} min_{b \in S} d(a,b) >
\epsilon$.

In particular, we allow only sampled comparisons $x :=_{\epsilon}
c$, which is a shorthand for $ \lfloor \frac {c}{\epsilon} \rfloor
\times \epsilon < x < \lceil \frac {c}{\epsilon} \rceil \times
\epsilon$.

\end{itemize}

\end{definition}

\begin{definition}

{$\epsilon$  Robust Hybrid Simulation Trajectory}: Given an initial
valuation of X i.e. $Val_0(X) = (x_0 = x0 ,x_1 = x1,\dots x_n=xn)$
and a path $\rho = \{ v_0, v_1 \dots v_m \}$ in a general hybrid
system $H = (X,V,E,v_i,\alpha,\beta)$, then $\tau (x_0,x_1, \dots
x_n) = f(t)$ is said to be an $\epsilon - $ robust hybrid simulation
trajectory \textit{iff}

\begin{itemize}

\item $\tau(0) = Val_0(X)$ and $Val_0(X) \in \alpha(v_0) $

\item if $x_k := e \in \psi(v_i,v_{i+1})$ then $ \tau(x_k,\sum_0^i (t_i) + = e )
$ else $\tau  (x_k, \sum_0^i (t_i) + ) = \tau (x_k,\sum_0^i(t_i)-)$

\item \textbf{Before executing the jump $(v_i,v_{i+1})$, $\tau  (x_k , \sum_0^i (t_i)
-)$ satisfies every precondition in $\phi(v_i,v_{i+1})$} $\epsilon-$
robustly .

\begin{itemize}

\item A linear constraint $c$ is $\epsilon-$ robustly satisfied by $X =
(x_0, x_1, \dots )$ iff for every $X'$ such that $d(X,X') \leq
\epsilon) $, $c(X')$ is true.

\end{itemize}

\item Within each location $v_i$ where the timed path has spent time $t_i$,

\begin{itemize}

\item $\forall t_i < t < t_{i+1},  \tau(v,\sum_0^i (t_i) + t)  $ is an
$\epsilon-$ simulation trajectory for location $v_i$ with initial
valuation $Val_0 = \tau( \sum_0^i (t_i) ) $.

\end{itemize}

\end{itemize}

\end{definition}

\begin{theorem}

If $\tau_{(v_0,x_0,x_1, \dots x_n)}$ be the $\epsilon-$ Robust
Hybrid Simulation Trajectory starting from the initial valuation
$Val_0 = \gamma_0 (x_0), \gamma_0 (x_1) \dots \gamma_0 (x_n)$, and
$H$ be a $\epsilon-$ structurally robust hybrid system, then
$tr_{concrete}$ corresponds to a \textbf{real} counterexample for
the hybrid system $H$.

\end{theorem}

\begin{proof}

The proof follows from the definition of $\epsilon-$ robust hybrid
automata and the notion of $\epsilon-$ hybrid simulation trajectory.

\end{proof}

\subsection{Simulation Based Abstraction Refinement}

Consider the \textit{concretization of a trace} $tr_{concrete}$ with
respect to the general hybrid automata $H$ obtained from a
\textit{trace} $tr$ of the abstract linear hybrid automata $H_a$.
Also, consider the \textit{guided hybrid simulation trajectory}
$\tau_{tr_{{concrete}}}$ corresponding to the concretization of the
trace $tr$ with respect to the general hybrid automata $H$.

\subsubsection{Metrics for distance between \textit{trace} and
\textit{trajectory}}

We define two distance metrics between a trace and the corresponding
guided hybrid simulation trajectory.

\begin{itemize}

\item $ D(t) =  d (\tau_{tr_{concrete}}(t), tr_{concrete}(t))$. \\ This
is simply a distance metric between corresponding points on the
trace and the trajectory. The metric $d$ may be the Euclidean
distance metric or the Manhattan distance metric (linear function).

\item $ D'(t) =  d' ( d ( \tau_{tr_{concrete}}(t),
tr_{concrete}(t)) , d ( \tau_{tr_{concrete}}(t-),
tr_{concrete}(t-))$ \\ This metric measures how rapidly the guided
hybrid simulation trajectory is moving away from the trace. The
metric $d$ may be the Euclidean distance metric or the Manhattan
distance metric (linear function), while the metric $d'$ may be the
real difference. $t-$ represents the last instant of time for which
the value of the concretized trace is known.

\end{itemize}

\subsubsection{Strategies to choose the location to be refined}

Let $t_i$ be the discrete point on the \textit{concretization of a
trace} i.e. on $tr_{concrete}$ for which $tr_{concrete}(t)$ is known
from the solution of the LP problem. There are few different
strategies to choose the location in the approximate linear hybrid
automata, where one needs to refine the abstract hybrid automata
$H_a$.

\begin{itemize}

\item $ min_i | D(t_i) | > \epsilon $ , where $\epsilon$ is an
empirically determined constant.

\item $  min_i |D'(t_i) - D' (t_{i-1})| > \epsilon $, where $\epsilon$ is an
empirically determined constant.

\item $  min_i |D'(t_i) / D' (t_{i-1})| > \epsilon $, where $\epsilon$ is an
empirically determined constant.

\end{itemize}

After finding out the point $t_i$ where one needs to refines the
location, the location $v_i$ at the time $t_i$ which needs to be
split is easily known from the \textit{concretized trace}.

\subsubsection{Choosing the variable to split the location}

When a simulation trajectory differs substantially from the trace
obtained by the LP solution, we need to split the location at which
the difference is substantial along a hyperplane such that the
abstract hybrid automata formed by the linear hybrid automata has a
trace that is close to the simulation trajectory. Let D be the
metric used to decide if a given location should be refined; then we
split those variables into half-spaces which have contributed beyond
a threshold to D.

\section{Conclusion and Future Work}
This early draft discusses the core issues involved in building a
CEGAR framework for analyzing nonlinear hybrid systems. The central
idea is to use linear programming as a mechanism for obtaining
feasible traces of the over-approximate linear hybrid automata (LHA)
abstractions and numerical simulation for obtaining a corresponding
trace of the original (possibly nonlinear) hybrid system. The
distance between these two traces is then used to guide the
refinement step in our CEGAR loop.

Several practical issues like the choice of the distance metrics,
the choice of picking up a particular solution to the linear program
and a characterization of the nonlinear functions which can be
handled using this paradigm have been left to a more complete
version of this draft. The techniques presented here are also being
implemented into a tool which will be a successor to the IRA
meta-tool for analyzing LHAs.

\section{Acknowledgement}
The authors thank Xuandong Li for introducing them to the use of
linear programming as an analysis technique for LHAs. Much of the
future work benefits from discussions with Ed Clarke and Bruce
Krogh.

\bibliography{nlha}

\end{document}